\documentclass[12pt,a4paper]{article}
\usepackage{amsmath,amssymb,amstext,amsthm}
\usepackage[utf8]{inputenc}
\usepackage{natbib}
\usepackage{graphicx}
\usepackage{bm}
\usepackage{color}
\usepackage{float}
\usepackage{algorithm}
\usepackage{CJK}
\usepackage{algpseudocode}
\usepackage{amsmath}
\usepackage{graphics}
\usepackage{epsfig}
\usepackage{adjustbox}
\usepackage{threeparttablex}
\usepackage{booktabs}
\usepackage{multirow}
\usepackage{subfigure}
\usepackage{diagbox}

\usepackage{url}

\usepackage[textwidth=14.5cm]{geometry}
\usepackage{blindtext}
\usepackage{enumitem}

\usepackage{geometry}
\usepackage{setspace}
\newtheorem{definition}{Definition}

\newcommand{\bea}{\begin{eqnarray*}}
\newcommand{\eea}{\end{eqnarray*}}
\newcommand{\Bea}{\begin{eqnarray}}
\newcommand{\Eea}{\end{eqnarray}}

\newcommand{\bfm}[1]{\mbox{\boldmath$#1$}}

\newcommand{\x}{\bfm{x}}
\newcommand{\U}{\bfm{U}}
\newcommand{\D}{\bfm{D}}

\newcommand{\Pcal}{\mathcal{P}}

\newcommand{\Tcal}{\mathcal{T}}
\newcommand{\Ucal}{\mathcal{U}}
\begin{document}
\title{\Large\bf  Construction of Uniform Designs over Continuous Domain in Computer Experiments}
\author{\large Jianfa Lai$^{1}$ , Kai-Tai Fang$^{2,3}$, Xiaoling Peng$^{2}$, and Yuxuan Lin$^{*,2}$}

\footnotetext[1]{Department of Mathematics, Hong Kong Baptist University, Kowloon Tong, Hong Kong}

\footnotetext[2]{Division of Science and Technology, BNU-HKBU United International College, Zhuhai 519085, China}
\footnotetext[3]{The Key Lab of Random Complex Structures and Data Analysis, The Chinese Academy of Sciences, Beijing, China}

\date{}
\maketitle
\begin{abstract}
Construction of uniform designs (UDs) has received much attention in computer experiments over the past decades, but most of previous works obtain uniform designs over a U-type by lattice domain. Due to increasing demands for continuous factors, UDs over continuous domain is in lack. Moreover, the uniformity can be further improved over a continuous domain. In this paper, we use coordinate descent methods with an initialization derived by threshold accepting (TA) algorithm to construct UDs over continuous domain with Centered $L_2$-discrepancy. The new UDs perform better than the recorded ones in several computer experiments  by the Kriging modeling.

\end{abstract}
\

\section{Introduction}
With the development of science and technology, computer experiments have gained more and more attention in recent years. Engineers and scientists implement computer simulations on physical systems due to the complex relationships between the inputs and outputs. \cite{fang2005design} gave a comprehensive introduction to the design and modeling of computer experiments.
In computer experiments, the true underlying model
\Bea\label{true-model}
Y=h(X_1,\ldots,X_s),\ \  (X_1,\ldots,X_s)\in \Tcal
\Eea
is known and usually too complicated to conduct some computation demanding works such as sensitivity analysis and optimization, where the response $Y$ depends on the factors $X_1,\ldots,X_s$ in the experimental domain $\Tcal$.
For a given input  $\{\x_1,\ldots,\x_n\} \in \Tcal$ representing $n$ experimental runs, one can obtain their output $\{y_i, i=1,\ldots, n\}$ from the true model (\ref{true-model}). Based on the observations of the
data set $\{y_i, \x_i, i=1, \ldots,n\}$, an approximate model 
\Bea\label{appro-model}
Y=\hat{h}(X_1,\ldots,X_s),
\Eea
called {\it metamodel} is built and used for some applications such as  optimization and sensitivity analysis. There are various modeling techniques  such as Kriging models, (orthogonal) polynomial regress models, Bayesian methods and neural networks. More details can be found in \cite{fang2005design}.

There are two key issues in computer experiments: design and modeling. The so-called space filling design has been recommended
for computer experiments.
The Latin hypercube sampling (LHS) proposed by \cite{mckay1979rj}  is one of the most popular space
filling design. With design points
 uniformly scattered on the experimental domain, the uniform design proposed by \cite{fang1980experimental}, \cite{wang1981note}
is another widely used space-filling design. 

The Latin hypercube sampling and uniform design are based on the estimation of overall mean of $Y$: 

\bea
E(Y)=\int_{\Tcal}h(\x)d\x=E(h(\x)),
\eea
where $\x=(x_1, \ldots,x_s)$ follows the uniform distribution on $\Tcal$.  Then $E(Y)$ can be estimated by the mean
$\bar{h}=\frac 1n\sum_{i=1}^n h(\x_i)$. LHS provides a more efficient estimate
of $E(Y)$ than the estimate by the simple random sampling. The Koksma-Hlawka inequality (see \cite{hua1981applications})
 gives the upper error bounds of the estimate as
\Bea\label{KH-ineq}
|E(h(\x))-\bar{h}|\le D(\Pcal)V(h),
\Eea
where $V(h)$ is a measure of the variation of $h$, and $D(\Pcal)$
is the star-discrepancy of $\Pcal$, a measure of the uniformity of $\Pcal$. The definition of $V(h)$ is in the sense of   Hardy and Krause (\cite{hua1981applications}).
Note that $V(h)$
is independent of the design points. Thus, given a bounded
$V(h)$, inequality (\ref{KH-ineq}) indicates that the more uniform a set
$\Pcal$ of points is over the experimental region $\Tcal$, the more
accurate $h$ is as an estimator of $E(h(\x))$. The uniform design is based on this overall mean model.

When the factors are quantitative and their domain is a rectangle $ a_i\le X_i\le b_i, i=1,\ldots, s$,
without loss of generality, we can assume the experimental domain is a unit cube $C^s=[0,1]^S$.
To measure the uniformity, \cite{hickernell1998generalized} proposed some new discrepancies such as
\textit{Centered $L_2$-discrepancy} (CD) and \textit{wrap-around $L_2$-discrepancy} (WD)
 to replace the star-discrepancy. \cite{zhou2013mixture} also suggested \textit{mixture discrepancy} (MD).
In this paper, we adopt CD. 
The squared CD has a computational formula
\begin{align}\label{CD-formula}
&CD^2(\Pcal)=(\frac{13}{12})^s-\frac{2}{n}\sum_{i=1}^{n}\prod_{k=1}^{s}(1+\frac{1}{2}|x_{ik}-0.5|-\frac{1}{2}|x_{ik}-0.5|^2)\nonumber\\
&+\frac{1}{n^2}\sum_{i=1}^{n}\sum_{j=1}^{n}\prod_{k=1}^{s}(1+\frac{1}{2}|x_{ik}-0.5|+\frac{1}{2}|x_{jk}-0.5|-\frac{1}{2}|x_{ik}-x_{jk}|),
\end{align}
where $\x_i=(x_{i1},\ldots,x_{is})$ is from a unit hypercube $C^s=[0,1]^s$. The uniform design is defined as follows.

	\begin{definition}
	For an experiment consists of $n$ runs relating to $s$ factors on an experimental region $\Tcal$, denote the experimental domain by $\Ucal$
	that is collection of sets $\Pcal=\{\x_1,\ldots,\x_n\} \in \Tcal$. Then a design, $\Pcal^*=\{\x_1^*,\ldots,\x_n^*\} \in \Tcal$, is
	called a uniform design under a given measure discrepancy $\mathcal{D}$ if
	\begin{equation}
	\label{UD}
	\mathcal{D}(\Pcal^*)=min_{\Pcal \in \Ucal}\mathcal{D}(\Pcal).
	\end{equation}
\end{definition}

Constructing a uniform design is an  NP hard problem that there is no exact algorithm for global optima within polynomial time. Besides, in practical experiments,
the feasible values of factors are limited so that the corresponding levels in designs are required to be discrete.
Therefore, for construction of uniform designs, traditional methods often focused on a lattice domain $\Ucal$ instead of the full set $\Pcal$. The U-type domain is prevalently accepted in existing literatures.
\begin{definition}\label{Utype}
Suppose a design $\U$ is an $n \times s$ matrix. Then it is said to be a q-level U-type design denoted as $\U(n,q^s)$ if in each column of $\U$,
the entries, $\frac{l-0.5}{q}$ for $l=1,\ldots,q$, appear equally often.
\end{definition}

Most of authors restrict a U-type design in an experimental region, $[0,1]^s$, defined in Definition \ref{Utype}.
The uniform design obtained from a U-type domain is called U-type uniform design, denoted as $\U_n(q^s)$.
There are many construction approaches to $\U_n(q^s)$, such as the good lattice point method,
the combinatorial construction method, cutting method, and
the foldover technique. Stochastic optimization algorithm is also feasible for this problem.  More details can refer to \cite{fang2018theory}.

The threshold accepting (TA) algorithm proposed by \cite{dueck1990threshold}, is a variation of the more widely used simulated annealing (SA) algorithm, that proposed by \cite{kirkpatrick1984optimization} through the emulation of the physical annealing process in solids. 
However, since SA algorithm involves too many parameters for the user choosing, its process convergence is slow.
Hence in many cases, TA as a descendent
has better performance than SA since it provides quicker convergence and better results during the same computational time. Multivariate or combinatorial optimization problems, including some NP-complete or NP-hard problems, have been tackled through TA, such as portfolio optimization problem (\cite{dueck1992new}),
NP-complete problem of optimal aggregation (\cite{chipman1995optimal}) and NP-hard traveling salesman problem (\cite{winker1994tuning}).

Referred to the definition of a uniform design in \eqref{UD}, generating a UD is a large scale integer programming problem with multimodal objective function ($i.e.$ discrepancies) over a discrete, large and complicated domain ($i.e.$ experimental region).
 \cite{winker1997application} and \cite{winker1998optimal}
firstly applied the TA to the calculation of the star discrepancy and construction of uniform designs on the lattice domain.
\cite{fang2017construction} discussed how to apply TA to construction of orthogonal designs. Most uniform designs recorded on the website
https://dst.uic.edu.cn/en/isci/uniform-design/uniform-design-tables have been obtained by TA algorithm.

Computer simulations play a dominant role in modern applications. Traditional computer experimental designs are in a lattice domain, in which the factors of interest are limited with discrete levels in experiments. However, with the improvement
of computational ability, factors in practical may refer to a continuous domain, that leads to a great loss of information if the experimenter still choose a design from a lattice domain. The uniform designs over continuous domain are suitable for this case, and easily to be carried out.
They should have better uniformity than U-type UDs as well, and as  consequence, perform better in modeling. Hence, construction of uniform designs over continuous domain is in badly need. Our paper is to propose an approach to obtaining the UDs over continuous domain, especially in a rectangle. 
Finding a uniform design in a continuous   $C^s=[0,1]^s$ is a high dimensional non-convex optimization problems. Here we only consider the case in which $n=q$ in Definition \ref{Utype}.
For a point set  $\Pcal=\{\x_1\ldots,\x_n\}=(x_{ik})_{n \times s}$ from  $C^s=[0,1]^s$, the objective is to  minimize the square of CD (\ref{CD-formula}):
\begin{equation}\label{UD-CD}
X=\mathop{\arg\min}_{\Pcal} CD^2(\Pcal).
\end{equation}

Gradient descent (\cite{ruder2016overview}) is a good way to solve such problems with a suitable initialization.
However, it costs too much for large sized designs
(i.e. $n$ and $s$ are large), which are prevalent in computer experiments. Coordinate descent (\cite{wright2015coordinate}) is also feasible for this problem.
Compared with gradient descent, the coordinate update is simpler and cheaper, especially for the high dimensional optimization problems.
Choosing a suitable initial design is very important, in this paper we adopt initial designs obtained from TA.  An introduction of TA  algorithm is given
in Section \ref{UDTA}.
In Section \ref{descent}, the implementation on constructing uniform designs through coordinate descent methods are conducted, in which iteration formula of CD is feasible to facilitate the efficiency of algorithm. To evaluate the modeling performance of new uniform designs over continuous domain, we adopt several case studies of computer experiments in Section \ref{cases}.

\section{Constructing Uniform Designs through TA Algorithm}\label{UDTA}
\begin{figure} [h]
\centering
\includegraphics[height=0.8\textwidth,width=0.8\textwidth]{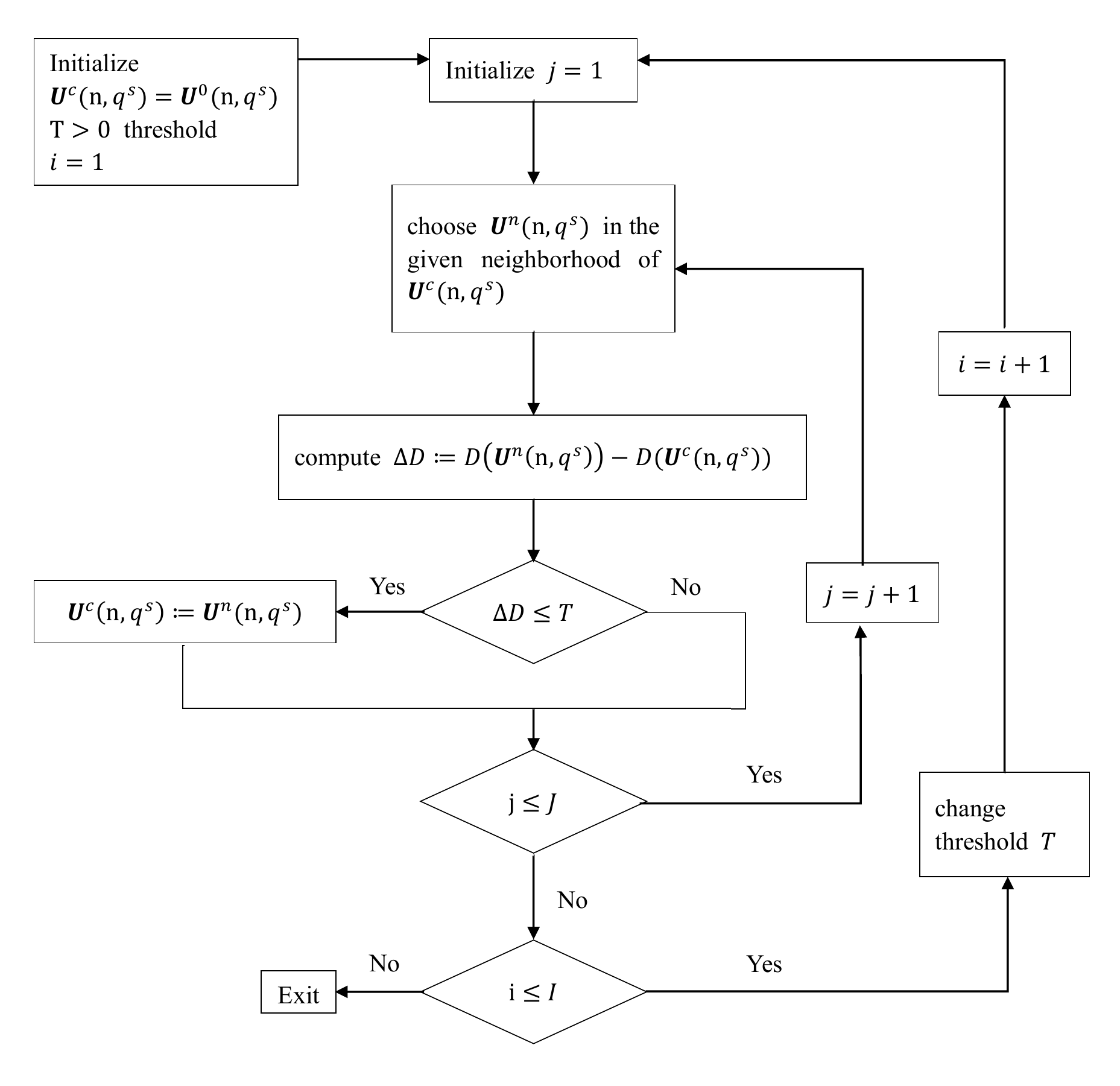}
\caption Threshold Accepting Algorithm for Generating Uniform Designs \label{TAFlowChart}
\end{figure}

The TA algorithm starts with an arbitrarily generated U-type design, say $\U^0(n,q^s)$,
and then it performs a large number of iterations. During each iteration, a new solution is generated and used for substituting
the current solution, say $\U^c(n,q^s)$, while in the first iteration $\U^c(n,q^s)=\U^0(n,q^s)$. The new solution $\U^{new}(n,q^s)$
is derived from a so-called \textit{neighborhood} of the current solution $\U^c(n,q^s)$, which is often defined as
a small permutation of the current solution. A decision rule in each iteration is to determine whether to accept the new solution and
update the current objective function value or not. We present the flow chart of TA algorithm for generating uniform designs
in Figure \ref{TAFlowChart}.

In uniform design construction, the purpose is to minimize the discrepancy. Hence, in each iteration, the results of the new candidate
and the current solution are compared through  $\Delta D=D(\U^c(n,q^s))-D(\U^{new}(n,q^s))$. A trivial local search algorithm accepts
$\U^{new}(n,q^s)$ if and only if $\Delta D\ge 0$. 
In this algorithm, the resulting solution
has a large probability that gets stuck in a bad local optimum especially as the objective function is multimodal. 
\cite{winker1998optimal} indicated that in the application to the traveling salesman problem, the result quality
of the trivial local search algorithm got stuck in a bad level and could not be improved by further increasing the number
of iterations. Therefore, TA adopts another acceptance criterion that allows certain temporary worsening with respect to
a predetermined threshold value, say $T\ge 0$. That implies in each iteration, a new candidate is accepted if and only if $\Delta D\leq T$.
Thus TA is also regarded as the refined local search algorithm.

The threshold value is not a constant but a finite sequence that is positive and decreasing to zero. As the threshold value is nonzero
in most of the iterations, even if the current solution is stuck in a bad local optimum, it has chance to leave. At the very end, the threshold
value ends up at zero, that ensures the solution of TA algorithm is still a local optimum. But the local optimum of TA has high quality
and at least is approaching to the global optimum.

\begin{figure} [h]
\centering
\includegraphics[height=0.4\textwidth,width=0.6\textwidth]{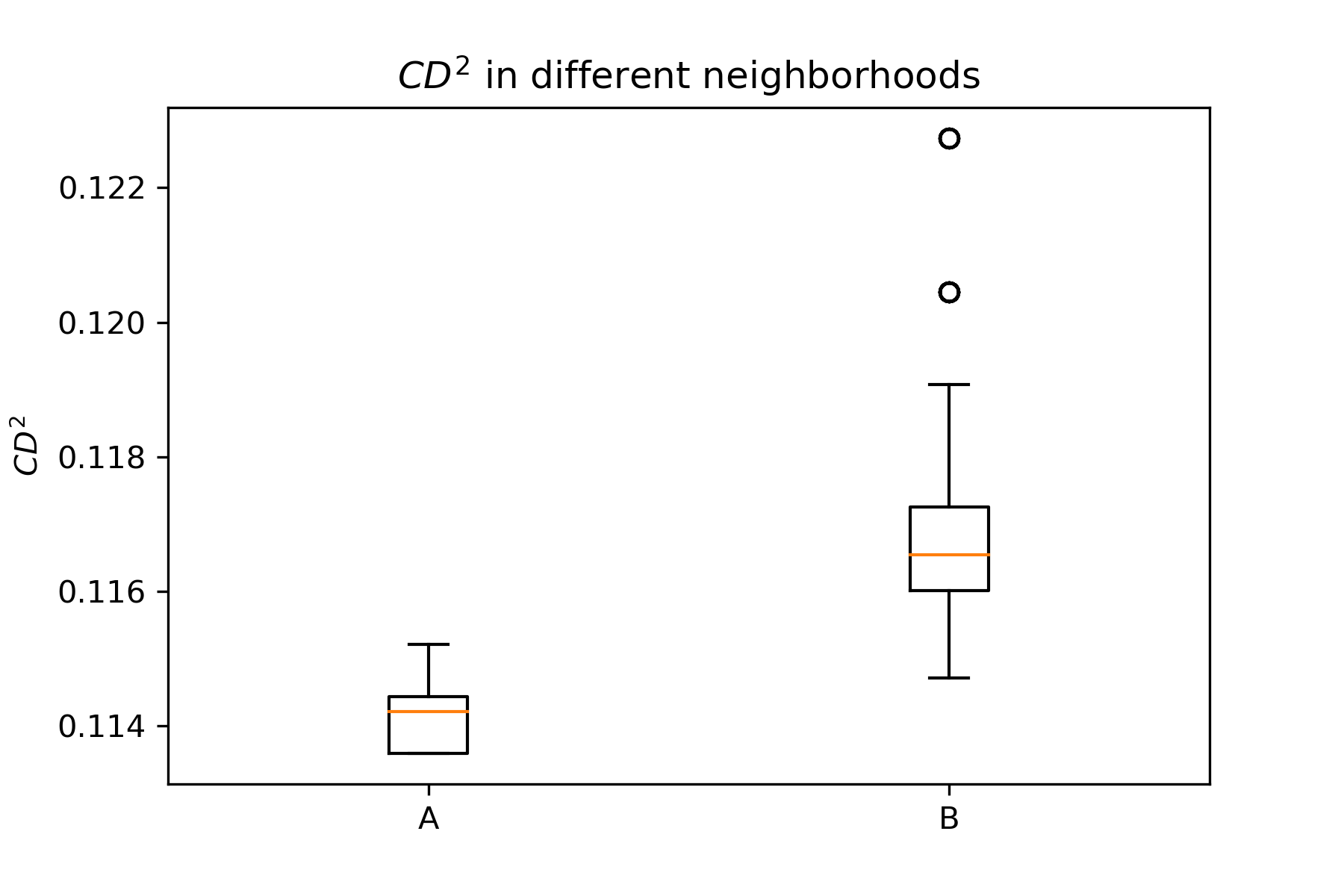}
\caption{The CD distributions in different neighborhoods of $U_{18}(3^7)$. \protect\\A: The box-plot of $CD^2$ in level permutation of one randomly chosen column of $U_{18}(3^7)$. \protect\\B: The box-plot of $CD^2$ in level-permutation neighborhood}
\label{CDdist}
\end{figure}

The key of TA algorithm is to define the neighborhood and threshold sequence. A U-type design is regarded as a neighborhood
of  the current design if they differ by elements in at most $k \leq s$ columns. Many authors set $k=1$,
and define each neighborhood by randomly exchanging two elements in a randomly chosen column of the current design (so-called TA neighborhood).
The definition of the neighborhoods provides an endogenous data-driven approach to generating a threshold sequence. In this paper, we adopt a recursive formula
$$T_i=\frac{I-i}{I}\times T_{i-1}$$ to derive the threshold sequence in an exponentially decreasing manner,  refer to \cite{fang2017construction}.
With the given parameters $\alpha \in (0,1)$, $I$ and $J$, the procedure of generating a threshold sequence is as follows:
 \begin{enumerate}[label=(\arabic*)]
\item generate the initial design of the TA algorithm, $\U^0(n,q^s)$;
\item find $J$ neighborhoods of the initial design, $\U^j(n,q^s)$, $j=1,\dots, J$;
\item calculate the squared CD values for each neighborhood, $CD^2(\U^j(n,q^s))$;
\item compute the range of the squared CD values of the $J$ neighborhoods, 
\\$range(CD^2(\U^j(n,q^s)))$;
\item set threshold basis $T_0=range(D(\U^j(n,q^s))) \times \alpha$ and $T_1=\frac{I-1}{I}\times T_0$;
\item generate the rest threshold sequence through $T_i=\frac{I-i}{I}\times T_{i-1}$, $i=2,\ldots,I$.
\end{enumerate}
However, the definition of  TA neighborhood is for the convenience of computation.
It is different from the mathematical neighborhood 
with respect to CD function. For instance, considering the uniform design $U_{18}(3^7)$ recorded on the website mentioned before,
Figure \ref{CDdist} gives the distributions of CD values of designs in different neighborhoods.
The right box-plot in Figure \ref{CDdist} plots the CD values of TA neighborhoods of $U_{18}(3^7)$. The left box-plot is for the designs
derived by level permutation on randomly chosen one column of $U_{18}(3^7)$. The CD value of TA neighborhoods may vary from
each other in a great amount, compared to level permuted designs. Hence, this definition of neighborhood is not feasible for continuous uniform designs.

\section{Coordinate Descent Algorithms with TA Initialization}\label{descent}

 Coordinate descent algorithms are a class of large-scale optimization algorithms that are feasible for various objective functions including non-smooth and non-convex ones. They have been successfully implemented on the problems in various modern applications such as machine learning and large-scale computational statistics. For some structured problems, coordinate descent algorithms have shown advantage on computation speed and convergence proof over traditional algorithms, such as gradient descent (GD), refer to \cite{shi2016primer}. A comprehensive introduction with applications to coordinate descent variants has been elaborated by \cite{wright2015coordinate} and \cite{shi2016primer}.  In this paper, three coordinate descent algorithms are introduced and discussed on their performance in constructing UDs.

Recall in \eqref{UD-CD} that constructing a uniform design under CD is an optimization problem with non-convex objective function,
\bea
y=\mathop{\arg\min}_{\Pcal=[0,1]^{ns}}g(x_1,\ldots,x_p),
\eea
where the objective function $g(x_1,\ldots,x_p)$ indicating CD in \eqref{CD-formula} has the second derivative, and $p=ns$. In traditional gradient optimization algorithm, there are two possible actions in each iteration.
Coordinate descent is a simple but efficient non-gradient optimization algorithm. The gradient optimization algorithm seeks for the minimum value of the function along the direction of the most rapid descent. However, coordinate descent minimizes the value of the objective function along the coordinate gradient in each step.

\subsection{\textbf{ \emph{ Coordinate Gradient Descent (CGD)}}}
\begin{figure}[h]
\centering
\subfigure{
\includegraphics[width=6cm]{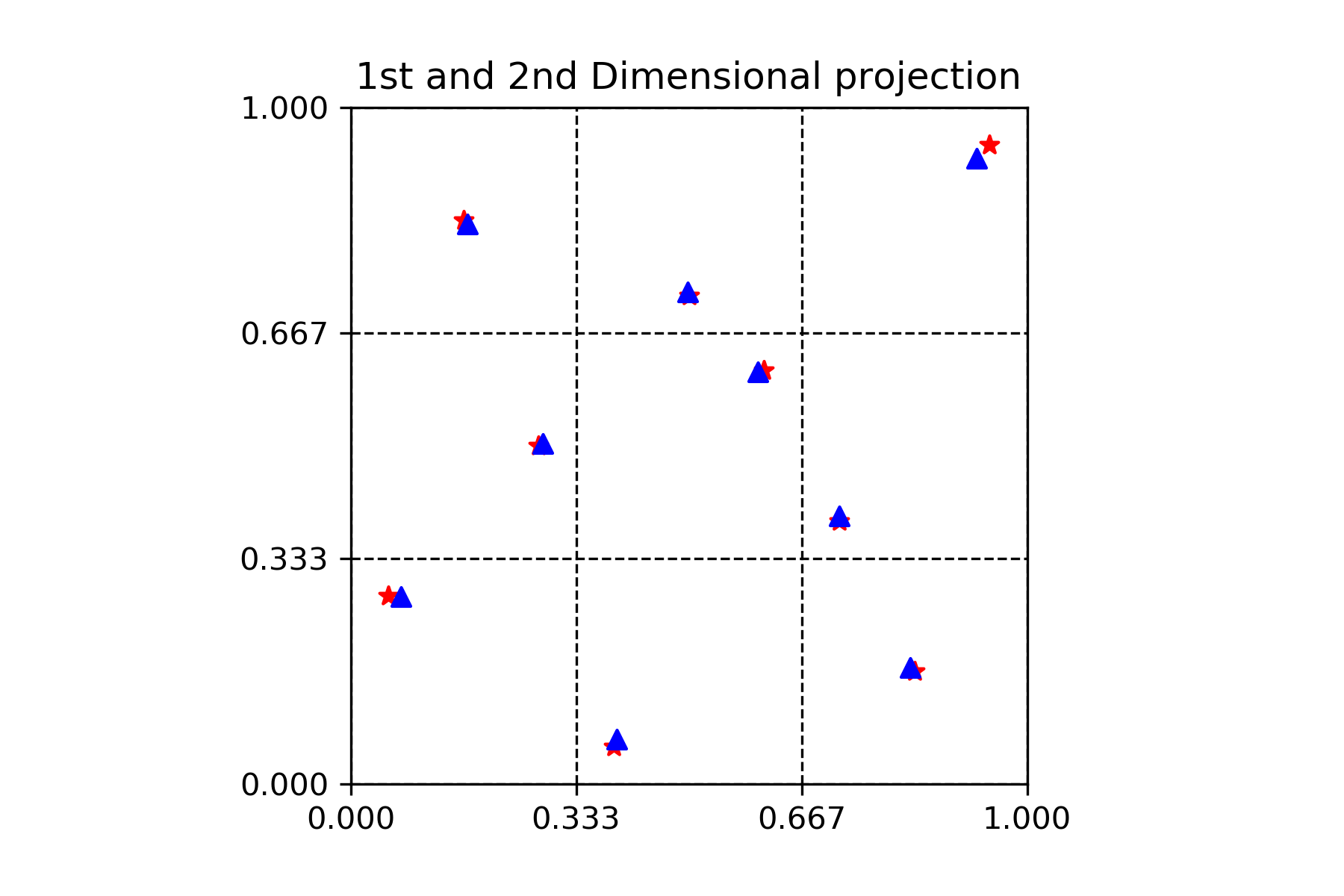}
}\hspace{-19mm}
\quad
\subfigure{
\includegraphics[width=6cm]{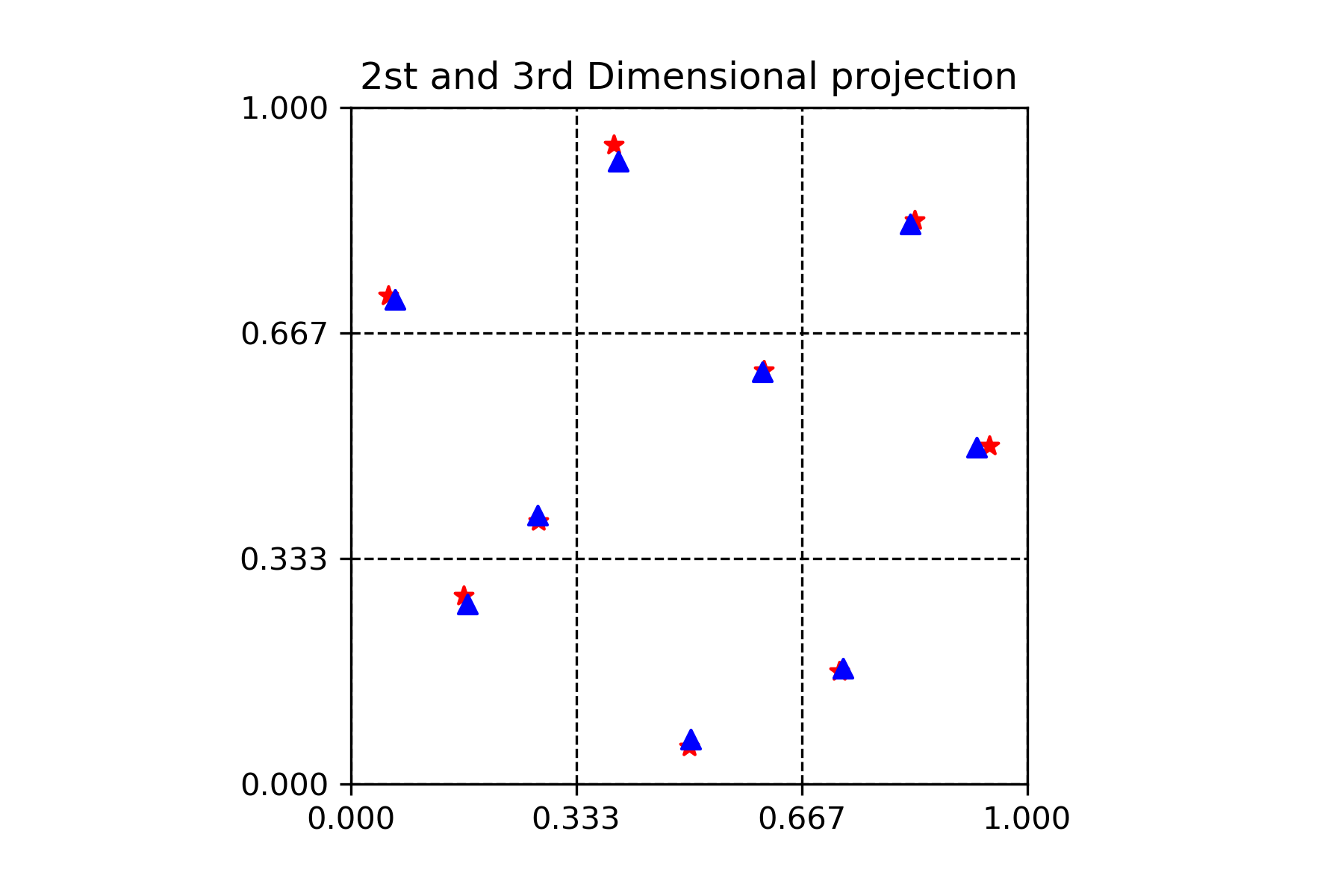}
}
\quad
\subfigure{
\includegraphics[width=6cm]{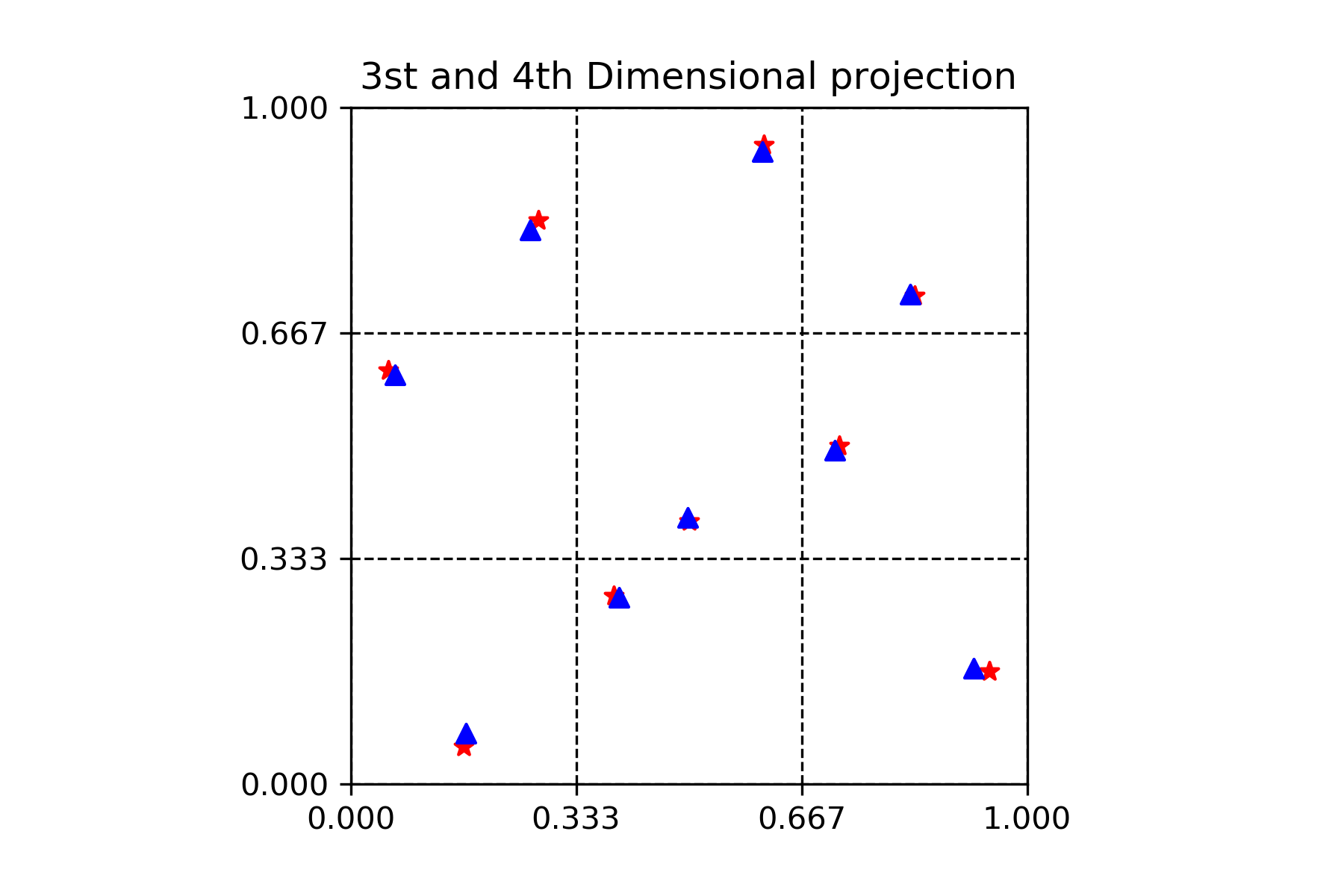}
}\hspace{-19mm}
\quad
\subfigure{
\includegraphics[width=6cm]{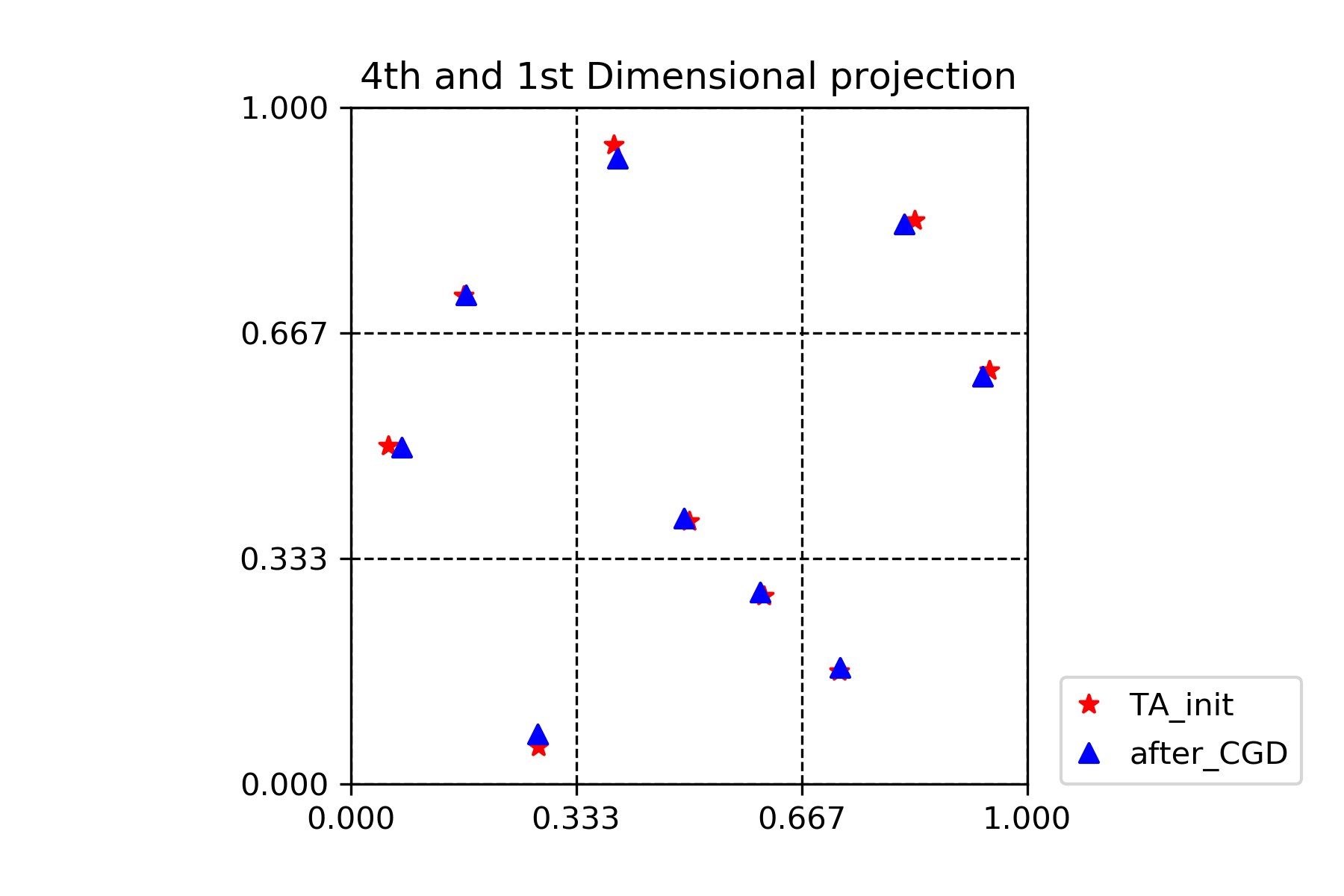}
}
\caption{Design points from any two-dimensional projection points}
\label{view2}
\end{figure}

To construct a UD, $U_n(q^s)$,  the optimization problem can be regarded as an $s$-dimensional problem with $n$ points on $[0,1]^s$. An $n \times s$ uniform design matrix can be viewed as $n$ design points each having $s$ dimension. Coordinate gradient descent
is to adjust the position of each design point in each dimension so that the whole design point set can be more uniformly
distributed in the domain. Figure \ref{view1} visualizes the design points during the optimization process. Figure \ref{view2}
presents the design points of $U_9(9^4)$ from any two-dimensional projection points. The visualization figures may help us comprehend
the objective uniformity measure CD adopted in this paper. 
The figures indicate that 
the whole design points become centrally aggregated after coordinate descent.
The following algorithm is the basic coordinate gradient descent method.
\begin{algorithm}[h]
\label{Alg1}
\caption{Coordinate Gradient Descent(CGD)}
\begin{algorithmic}[1]
\State Given n,s
\State Initialization $X^0$ and choose $\epsilon$, the step size $\delta$ and the maximum epoch M.
\State Create cd list and matrix list
\For{each $m\in [0,M]$}
    \For{each $i\in [0,n]$}
        \For{each $j\in [0,s]$}

            \State Calculate the Coordinate Gradient descent, g=$\frac{dCD^2(X^m)}{dx^{m}_{ij}}$

            \State $x^{m+1}_{ij} = x^{m}_{ij} + g*\delta$
        \EndFor
    \EndFor
    \State if $|CD^2(X^{m+1})-CD^2(X^m)|<\epsilon$, then break the for-loop
\EndFor
\label{code:recentEnd}
\end{algorithmic}
\end{algorithm}

\begin{figure}[h]
\label{view1}
\centering
\includegraphics[height=0.5\textwidth,width=1\textwidth]{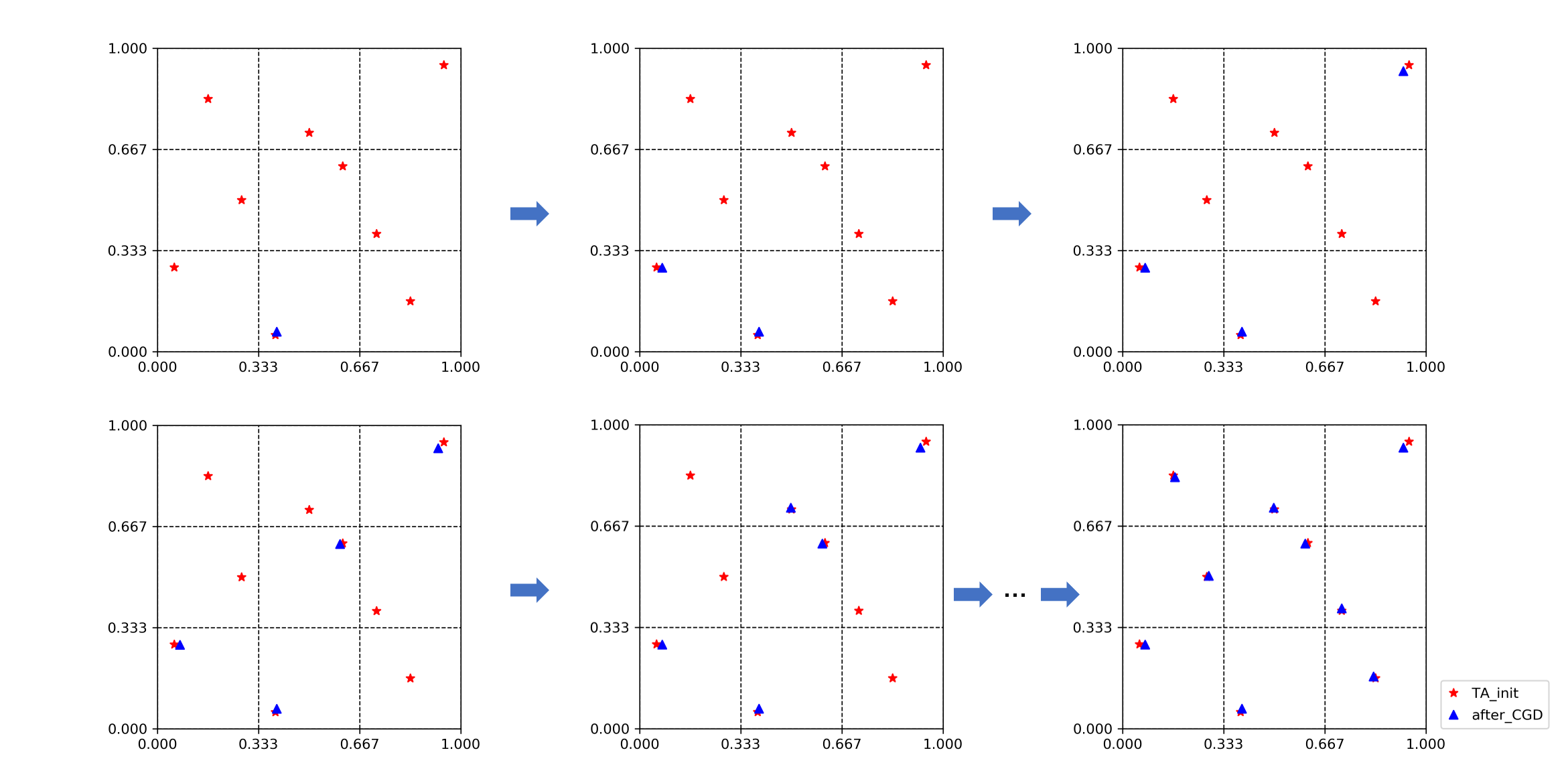}
\caption{Visualization of design points during the optimization process} \label{view1}
\end{figure}

\clearpage
\subsection{\textbf{ \emph{ Coordinate Zero-gradient (CZG)}}}
Since the coordinate gradient in Algorithm 1, $g=\frac{dCD^2(X^m)}{dx^{m}_{ij}}$, is known, we can set $g=0$ to find the coordinate value while fixing other coordinate values. 

\begin{align}\label{gradient}
&\frac{dCD^2(\Pcal)}{dx_{ij}}=-\frac{2}{n}\left \{\prod_{q\neq j}^s\left [1+\frac{1}{2}|x_{iq}-\frac{1}{2}|-\frac{1}{2}(x_{iq}-\frac{1}{2})^2 \right ] \right \} \nonumber\\
& \times \left [\frac{1}{2}sgn(x_{ij}-\frac{1}{2})-(x_{ij}-\frac{1}{2}) \right ]\nonumber\\
&+\frac{1}{n^2} \prod_{q\neq j}^s\frac{1}{2}(1+|x_{iq}-\frac{1}{2}|)  sgn(x_{ij}-\frac{1}{2})\nonumber\\
&+\frac{2}{n^2}\sum_{k\neq i}^n \Bigg \{\prod_{q\neq j}^s \left [1+\frac{1}{2}|x_{iq}-\frac{1}{2}|+ \frac{1}{2}|x_{kq}-\frac{1}{2}|-\frac{1}{2}|x_{iq}-x_{kq}| \right ]\nonumber\\
&\times \left [\frac{1}{2}sgn(x_{ij}-\frac{1}{2})-\frac{1}{2}sgn(x_{ij}-x_{kj})\right ] \Bigg\},
\end{align}
where the sgn function is:
\begin{equation}
sgn(x) =
    \left \{
            \begin{aligned}
                \; 1, & \quad & x < 0, \\
                \; 0, & \quad & x = 0, \\
                \; -1,  & \quad & x < 0.
        \end{aligned}
    \right.
\end{equation}
The coordinate value, $x_{ij}$, can be derived through taking $\frac{dCD^2(\Pcal)}{dx_{ij}}=0$. However, some of $x_{ij}$ in sgn function
can not be solved through explicit expression. Here we adopt recursion with an initialization, denoted as $x^{0}_{ij}$, and a maximum number
of iterations, $T$, while other coordinate values  $\{x_{kq}, k\neq i, q \neq j\}$, are fixed. The recursive formula is given as

\begin{equation}
 x^{t}_{ij} = \frac{A}{B} +\frac{1}{2},
\end{equation}
where
\begin{align}
&A=-\frac{2}{n}\left \{\prod_{q\neq j}^s\left [1+\frac{1}{2}|x_{iq}-\frac{1}{2}|-\frac{1}{2}(x_{iq}-\frac{1}{2})^2 \right ] \right \}\times \frac{1}{2}sgn(x^{t-1}_{ij}-\frac{1}{2})\nonumber\\
&+\frac{1}{n^2} \prod_{q\neq j}^s\frac{1}{2}(1+|x_{iq}-\frac{1}{2}|)  sgn(x^{t-1}_{ij}-\frac{1}{2})\nonumber\\
&+\frac{2}{n^2}\sum_{k\neq i}^n \Bigg \{\prod_{q\neq j}^s \left [1+\frac{1}{2}|x_{iq}-\frac{1}{2}|+ \frac{1}{2}|x_{kq}-\frac{1}{2}|-\frac{1}{2}|x_{iq}-x_{kq}| \right ]\nonumber\\
&\times \left [\frac{1}{2}sgn(x^{t-1}_{ij}-\frac{1}{2})-\frac{1}{2}sgn(x^{t-1}_{ij}-x_{kj})\right ] \Bigg\}
\end{align}
and
\begin{equation}
B = -\frac{2}{n}\prod_{q\neq j}^s\left [1+\frac{1}{2}|x_{iq}-\frac{1}{2}|-\frac{1}{2}(x_{iq}-\frac{1}{2})^2 \right ]
\end{equation}
for $t \in [0,T]$. Through plenty of experiments, $T=1$ is enough for getting good results.

\begin{algorithm}[h]
\label{Alg2}
\caption{Coordinate zero-gradient}
\begin{algorithmic}[1]
\State Given n,s
\State Initialization $X^0$ and choose $\epsilon$ and the maximum epoch M.
\State Create cd list and matrix list
\For{each $m\in [0,M]$}
    \For{each $i\in [0,n]$}
        \For{each $j\in [0,s]$}

            \State Let \{$x^m_{kl}$, $k\neq i$, $l \neq j$\} be fixed. Calculate $\hat{x}^m_{ij}$ such that Coordinate Gradient descent, $\frac{dCD^2(\Pcal)}{d\hat{x}^m_{ij}}=0$.

            \State $x^{m+1}_{ij} = \hat{x}^m_{ij}$
        \EndFor
    \EndFor
    \State if $|CD^2(X^{m+1})-CD^2(X^m)|<\epsilon$, then break the for-loop
\EndFor
\label{code:recentEnd}
\end{algorithmic}
\end{algorithm}
In our experiments, the coordinate zero-gradient method is advantaged in computation
speed. For the same optimization task, it takes several minutes through
coordinate gradient descent method but several seconds
through coordinate zero-gradient method. However, even though the
final optimization result is good enough, the CD values during the
optimization process is unstable. For instance, sometimes there may
be violent fluctuations of the CD values.

\subsection{\textbf{ \emph{ Coordinate Descent with Fixed Step Size (CDFSS)}}}

Considering the limit of minimum unit of variables in practical experiments, coordinate descent with fixed step size are suitable under such circumstances. The idea of this method is to replace the coordinate gradient with $\{-1,1\}$ and the step size can be set as the minimum unit of variables. In each step, the coordinate with the rapidest descent will be chosen. The algorithm will be stopped until there is no further descent in any coordinate.  

This method can be easily applied in the experiments with minimum unit due to the equipment or technology limitations. For example, one of the variable, temperature, in the industry experiment is from 0 to T. Due to the limitations of temperature detection and control technology, the minimum unit is 0.1 degree. With CDFSS, the step size of the variable can be set as $\frac{0.1}{T-0}$. Similarly for other variables, the step size can be set different values according to the corresponding limitations. Thus, the resulted uniform design is feasible under practical experimental requirements. 
\begin{algorithm}[h] 
\label{Alg3}
\caption{Coordinate Descent with Fixed Step Size}
\begin{algorithmic}[1] 

\State Given n,s 
\State Initialization $X^0$ and choose the step size in each column $\{\delta_j, j \leq s \}$ and the maximum epoch M.
\State Create CD list and matrix list
\For{each $m\in [0,M)$} 
    \For{each $i\in [0,n)$}
        \For{each $j\in [0,s)]$}
            \For{each $a\in \{-1,1\}$}
            
                \State $\hat{X}$ be the copy of $X^m$
            
                \State $\hat{x}_{ij} = \hat{x}_{ij} + a*\delta_j$
            
                \State Calculate $cd=CD^2(\hat{X})$ and put $cd$ into cd list and put $\hat{X}$ into matrix list 
            \EndFor
        \EndFor 
    \EndFor 
    \State Find the matrix with minimum CD in matrix list and if the CD is smaller than the CD of $X^m$, then let $X^{m+1}$ be such matrix, otherwise break the for-loop
\EndFor 
\label{code:recentEnd} 
\end{algorithmic} 
\end{algorithm}

Actually, in the experiments, these three algorithms may get different results with the same initialization. With the same initialization, the aforementioned three algorithms are conducted respectively. We present the CD values in Table \ref{CDcompare}. Table \ref{CDcompare} shows that the final uniform designs of these algorithms 
may be different due to the complexity of non-convex high dimensional optimization. But using better initialization ($U_{18}(18^7)$ on the website) facilitates the optimization process.
\begin{table}[h]
\centering
\caption{$CD^2$ of different algorithms with the same initialization}
\begin{tabular}{|l|c|r|} 
\hline 
\diagbox{Algorithm}{Initialization}  &  Website & $X_{ij} \sim U(0,1)$ \\
\hline  
GD & 0.033972  & 0.052300 \\
\hline 
CGD & 0.033972 & 0.041678 \\
\hline 
CZG & 0.033972 & 0.052690 \\
\hline 
CDFSS & 0.033972 & 0.042079 \\
\hline 
\end{tabular}
\label{CDcompare}
\end{table}

\subsection{\textbf{ \emph{ TA Initialization}}}
Due to the non-convexity of $CD^2$ in \eqref{CD-formula}, the initialization is significant in the gradient-based methods.
With different initialization, various local minimum will be achieved by using the same coordinate descent algorithm.
For example, when constructing $U^{new}_{18}(18^7)$ through different initial designs, coordinate gradient descent obtains different results shown in Table \ref{diffX0}.
\begin{table}[h]
\centering
\caption{Coordinate gradient descent with different initialization}
\begin{tabular}{|l|c|}
\hline
Initialization type
 & $CD^2$ \\
\hline
$X_{ij}\sim U(0,1)$  & 0.041678 \\
\hline
$X_{ij}\sim N(0.5,0.1^2)$ & 0.061949  \\
\hline
$X_{ij}=0$  & 0.041114  \\
\hline
$X_{ij}=1$ & 0.041227  \\
\hline
$U_{18}(18^7)$ on website & 0.033972  \\
\hline
\end{tabular}
\label{diffX0}
\end{table}

TA algorithm has successfully obtained many uniform designs over the U-type lattice point domain on the website mentioned above. Hence,
intuitively we adopt the uniform design derived by TA algorithm as the initialization. Then subsequently, we conduct the coordinate descent methods to optimize the uniform designs over a continuous domain, $[0,1]^s$. Table \ref{CDTA} presents
the CD values of uniform designs derived by TA over a U-type domain and by TA+CGD over a continuous domain. The CD values can be
further improved over a continuous domain through coordinate descent methods, compared to the existing UDs on the website.

\begin{table}[h]
\centering
\caption{CD values of uniform designs through TA and TA+CGD}
\begin{tabular}{|l|c|r|}
\hline
Algorithms & $U_{18}(18^7)$ & $U_{27}(27^{13})$ \\
\hline
TA  & 0.035403 & 0.228455 \\
\hline
TA + CGD & 0.033972 & 0.198073 \\
\hline
\end{tabular}
\label{CDTA}
\end{table}

\begin{table}[h]
   \centering
   \caption{$U^{new}_{18}(18^7)$ obtained by TA+CGD}
   \begin{tabular}{rrrrrrr}
   \hline
0.9080  & 0.5238  & 0.6906  & 0.8001  & 0.7509  & 0.3578  & 0.9071  \\
0.3050  & 0.5838  & 0.6351  & 0.0531  & 0.6986  & 0.7447  & 0.1393  \\
0.7475  & 0.2567  & 0.4738  & 0.2078  & 0.9052  & 0.8544  & 0.6415  \\
0.9457  & 0.1462  & 0.3609  & 0.4796  & 0.3579  & 0.6288  & 0.0542  \\
0.4704  & 0.6925  & 0.5757  & 0.9079  & 0.1977  & 0.2049  & 0.0972  \\
0.1438  & 0.4809  & 0.0946  & 0.2525  & 0.0549  & 0.5210  & 0.2979  \\
0.5264  & 0.0986  & 0.8501  & 0.1441  & 0.1533  & 0.4260  & 0.7990  \\
0.3570  & 0.0529  & 0.1464  & 0.5804  & 0.8045  & 0.2527  & 0.4264  \\
0.0957  & 0.7482  & 0.9468  & 0.5135  & 0.8528  & 0.5862  & 0.5889  \\
0.6366  & 0.4197  & 0.0528  & 0.9471  & 0.5825  & 0.6931  & 0.7531  \\
0.0546  & 0.3076  & 0.4149  & 0.6973  & 0.3062  & 0.0959  & 0.6922  \\
0.2009  & 0.1964  & 0.7458  & 0.8505  & 0.4748  & 0.9032  & 0.3564  \\
0.7971  & 0.8605  & 0.8011  & 0.6355  & 0.0960  & 0.7947  & 0.4694  \\
0.4159  & 0.6462  & 0.2524  & 0.4225  & 0.2570  & 0.9456  & 0.9456  \\
0.5711  & 0.9058  & 0.2994  & 0.7430  & 0.9456  & 0.4756  & 0.2040  \\
0.2435  & 0.9512  & 0.5211  & 0.3061  & 0.5330  & 0.3026  & 0.8508  \\
0.6876  & 0.3689  & 0.9043  & 0.3633  & 0.6321  & 0.0539  & 0.2502  \\
0.8534  & 0.7923  & 0.1928  & 0.0974  & 0.4286  & 0.1500  & 0.5260  \\
   \hline
   \end{tabular}
   \label{UDwood}
\end{table}

\begin{table}[h]\small
   \centering
   \caption{$U^{new}_{27}(27^{13})$ obtained by TA+CGD}
   \resizebox{\textwidth}{!}{
   \begin{tabular}{rrrrrrrrrrrrr}
   \hline
   0.0971  & 0.8438  & 0.2913  & 0.5290  & 0.7735  & 0.4409  & 0.8424  & 0.6708  & 0.4005  & 0.1238  & 0.3293  & 0.6793  & 0.8683  \\
0.5663  & 0.8711  & 0.7423  & 0.3992  & 0.1332  & 0.1908  & 0.1838  & 0.3580  & 0.5025  & 0.8424  & 0.5971  & 0.7083  & 0.0939  \\
0.7020  & 0.0962  & 0.2268  & 0.1853  & 0.5957  & 0.3299  & 0.5591  & 0.0915  & 0.1942  & 0.4693  & 0.6391  & 0.7802  & 0.7763  \\
0.1914  & 0.5124  & 0.6684  & 0.7741  & 0.5287  & 0.1272  & 0.2272  & 0.2965  & 0.3338  & 0.2203  & 0.8354  & 0.2634  & 0.8097  \\
0.2906  & 0.2188  & 0.4044  & 0.3498  & 0.6775  & 0.1570  & 0.9011  & 0.1962  & 0.8988  & 0.5647  & 0.4023  & 0.3885  & 0.1243  \\
0.4771  & 0.1907  & 0.8427  & 0.9289  & 0.7377  & 0.9084  & 0.0971  & 0.5278  & 0.2559  & 0.4013  & 0.4324  & 0.7423  & 0.5594  \\
0.3673  & 0.1224  & 0.8728  & 0.6715  & 0.3999  & 0.2237  & 0.6704  & 0.8715  & 0.6986  & 0.8150  & 0.2553  & 0.5226  & 0.7389  \\
0.8424  & 0.4354  & 0.9068  & 0.8085  & 0.9281  & 0.4679  & 0.7395  & 0.1242  & 0.4656  & 0.7069  & 0.5278  & 0.1253  & 0.4302  \\
0.0717  & 0.1596  & 0.2587  & 0.2915  & 0.2922  & 0.5324  & 0.3007  & 0.5610  & 0.5250  & 0.9289  & 0.8047  & 0.0925  & 0.5224  \\
0.2272  & 0.6788  & 0.5259  & 0.7056  & 0.0705  & 0.8810  & 0.6402  & 0.1554  & 0.6313  & 0.6011  & 0.9295  & 0.6398  & 0.6283  \\
0.9267  & 0.7367  & 0.8156  & 0.3238  & 0.2645  & 0.5657  & 0.3593  & 0.2191  & 0.1246  & 0.0966  & 0.3680  & 0.5100  & 0.6731  \\
0.9006  & 0.3594  & 0.1018  & 0.4724  & 0.8127  & 0.2550  & 0.1526  & 0.7021  & 0.6665  & 0.2897  & 0.9045  & 0.5622  & 0.3623  \\
0.6695  & 0.0708  & 0.4881  & 0.6381  & 0.5080  & 0.8414  & 0.4007  & 0.9288  & 0.4352  & 0.0711  & 0.6766  & 0.3326  & 0.0710  \\
0.5170  & 0.4010  & 0.4748  & 0.1214  & 0.0934  & 0.4096  & 0.0724  & 0.7373  & 0.2922  & 0.6378  & 0.1547  & 0.3624  & 0.9275  \\
0.1574  & 0.3288  & 0.6032  & 0.4884  & 0.8388  & 0.7119  & 0.4358  & 0.2642  & 0.0962  & 0.8977  & 0.0973  & 0.5941  & 0.2872  \\
0.6397  & 0.8051  & 0.1902  & 0.9022  & 0.7080  & 0.5996  & 0.4721  & 0.4047  & 0.8367  & 0.8721  & 0.7358  & 0.4241  & 0.9042  \\
0.8098  & 0.7682  & 0.6383  & 0.7497  & 0.5623  & 0.3728  & 0.3233  & 0.7705  & 0.9292  & 0.5323  & 0.1828  & 0.8078  & 0.2576  \\
0.4356  & 0.9014  & 0.7820  & 0.2190  & 0.6455  & 0.6712  & 0.8026  & 0.8402  & 0.2229  & 0.5137  & 0.8732  & 0.2930  & 0.3339  \\
0.5292  & 0.2573  & 0.1643  & 0.8350  & 0.2219  & 0.6356  & 0.7776  & 0.3306  & 0.5549  & 0.1957  & 0.1271  & 0.8760  & 0.3938  \\
0.3232  & 0.5606  & 0.3683  & 0.2547  & 0.9073  & 0.8088  & 0.2573  & 0.9027  & 0.7356  & 0.6824  & 0.5189  & 0.9025  & 0.6990  \\
0.1306  & 0.4634  & 0.9265  & 0.0734  & 0.3654  & 0.4859  & 0.5235  & 0.4446  & 0.8092  & 0.1592  & 0.7133  & 0.8388  & 0.2217  \\
0.4013  & 0.9279  & 0.3244  & 0.5670  & 0.4259  & 0.7460  & 0.1239  & 0.0721  & 0.7729  & 0.3316  & 0.2306  & 0.1871  & 0.4684  \\
0.8678  & 0.5964  & 0.0736  & 0.0967  & 0.4722  & 0.9262  & 0.7000  & 0.4762  & 0.3608  & 0.7794  & 0.2910  & 0.4719  & 0.1625  \\
0.7417  & 0.5304  & 0.4385  & 0.6016  & 0.3333  & 0.0932  & 0.9281  & 0.5975  & 0.0718  & 0.7427  & 0.7742  & 0.9281  & 0.5089  \\
0.7788  & 0.3003  & 0.7021  & 0.4343  & 0.1600  & 0.7792  & 0.8683  & 0.6321  & 0.8693  & 0.3659  & 0.5660  & 0.2166  & 0.8376  \\
0.2576  & 0.6397  & 0.1321  & 0.8754  & 0.1864  & 0.2929  & 0.5135  & 0.8070  & 0.1546  & 0.4325  & 0.4745  & 0.1554  & 0.1941  \\
0.6064  & 0.6963  & 0.5617  & 0.1565  & 0.8705  & 0.0711  & 0.5982  & 0.5135  & 0.5994  & 0.2536  & 0.0715  & 0.0715  & 0.5914  \\
      \hline
   \end{tabular}}
   \label{UDwood}
\end{table}

To further explore the performance of new uniform designs, we implement them on several computer experiments in the next section. In these case studies, we use Kriging modeling technique
to obtain an approximate model for responses and factors. We evaluate the mean squared errors (MSE) when predicting untried points
through the approximate models trained by the new uniform designs, compared with the ones recorded on the website, and the Latin hypercube sampling.

\section{Modeling Performance of Uniform designs over Continuous Domain}
\label{cases}
The question is requested whether the new uniform designs with smaller CD on the continuous domain has better performance in modeling.
We apply the new uniform designs in several computer experiments as case studies. There are many modeling techniques among which
we adopt Kriging modeling technique in this study.  \cite{fang2005design} gave an elaborate introduction to Kriging models.
\begin{definition}
Suppose that
$\x_i,\ i=1,\ldots,n$ are design points over an $s$-dimensional experimental domain, and $y_i=y(\x_i)$ is the corresponding output.
The universal Gaussian Kriging model is defined as
$$y(\x)=\sum_{j=0}^{L}\beta_jB_j(\x)+z(\x),$$
where  the set of $B_j$ is a chosen polynomial basis over the
design region and $z(\x)$ is a Gaussian process.
The ordinary Kriging model is the most commonly used in practice, defined as
$$y(\x)=\mu+z(x),$$
where $\mu$ is the overall mean of $y$.
\end{definition}
DACE packakge in MATLAB is employed for Kriging modeling. There are several candidate basis in this package, such as {\it Poly0}
representing ordinary Kriging models, {\it Poly1} representing first-order polynomial function as basis, and {\it Poly2} representing
second-order polynomial function as basis.
We evaluate the modeling performance through comparing the prediction MSE on untried $1000$ points of Kriging models trained
by the new uniform designs and the recorded ones in the website. The MSE is defined as follows.
$$MSE=\frac{1}{M}\sum^{M}_{m=1}(y_m-\hat{y}_m)^2.$$

Since Latin hypercube sampling (LHS) has been popular in computer experiments, we also provide the prediction MSE of
the Kriging models trained by LHS and mid-point LHS (MLHS), defined in Definition \ref{LHS}. Since each LHS or MLHS is a random sample,
we generate ten LHS or MLHS each time and conduct Kriging modeling respectively. Subsequently, we present the average prediction
MSE of each Kriging model trained by these ten LHS or MLHS, denoted as Avg. LHS or Avg. MLHS in Table \ref{MSEwood} and \ref{MSEcamel}.

\begin{definition}\label{LHD}
A Latin hypercube design (LHD) with $n$ runs and $s$ factors, denoted by $LHD(n,s)$, is an $n\times s$ matrix, in which each column
is a random permutation of {1,2,\ldots,n}.
\end{definition}
\begin{definition}\label{LHS}
Let $\pi_j(1),\ldots,\pi_j(n)$, $j=1, \ldots, s$, denote the $s$ permutations of $LHD(n,s)$. Take mutually independent $ns$ uniform
random variables $U_k^j\sim U(0,1)$, $k=1, \ldots, n$, $j=1, \ldots, s$. Let $\x_k=(x_k^1,\ldots, x_k^s)$,
where $x_k^j=\frac{\pi_j(k)-U_k^j}{n},$ $k=1, \ldots, n$, $j=1, \ldots, s$. Then $\D_n=\{\x_1,\ldots,\x_n\}$ is a LHS and denoted
by $LHS(n,s)$. If for each $x_k^j$, we fix at a mid-point instead of a random number, that is $x_k^j=\frac{\pi_j(k)-0.5}{n}$,
$\D_n$ is a mid-point LHS, denoted as MLHS.
\end{definition}

\subsection{\textbf{ \emph{ Wood Model}}}
A wood function is notorious due to its difficulty on optimization, defined by 
\begin{align*}
Y=&100(x_1^2-x_2)^2+(1-x_1)^2+90(x_4-x_3^2)^2+(1-x_3)^2\\
&+10.1((x_2-1)^2+(x_4-1)^2)+19.8(x_2-1)(x_4-1),
\end{align*}
for $(x_1,x_2,x_3,x_4) \in [-2,2]^4$. The global minimum is at $x^*=[1,1,1,1]$ with $Y^*=0$. We apply several new uniform designs over continuous domain, denoted as $U^{new}_{n}(n^s)$, and the corresponding recorded ones, $U_{n}(n^s)$ on the website respectively into wood model, in which 9-run and 16-run designs shown in Table \ref{UDwood}.
\begin{table}[h]
   \centering
   \caption{$U_{9}(9^4)$, $U^{new}_{9}(9^4)$, $U_{16}(16^4)$, and $U^{new}_{16}(16^4)$ in wood model}
   \resizebox{\textwidth}{!}{ 
   \begin{tabular}{llll|rrrr|rrrr|rrrr}
   \hline
      \multicolumn{4}{c|}{$U_{9}(9^4)$} & \multicolumn{4}{c|}{$U_{9}(9^4)\in [-2,2]^4$} & \multicolumn{4}{c|}{$U^{new}_{9}(9^4)\in [0,1]^4$} &\multicolumn{4}{c}{$U^{new}_{9}(9^4)\in [-2,2]^4$}\\
   \hline
   4 & 1 & 7 & 5 & -0.5 & -2 & 1 & 0 & 0.3942  & 0.0690  & 0.7114  & 0.5978  & -0.4232  & -1.7242  & 0.8457  & 0.3912  \\
1 & 3 & 4 & 3 & -2 & -1 & -0.5 & -1 & 0.0743  & 0.2762  & 0.3966  & 0.2761  & -1.7030  & -0.8953  & -0.4136  & -0.8957  \\
9 & 9 & 5 & 4 & 2 & 2 & 0 & -0.5 & 0.9303  & 0.7215  & 0.6060  & 0.3940  & 1.7214  & 0.8861  & 0.4240  & -0.4240  \\
6 & 6 & 6 & 9 & 0.5 & 0.5 & 0.5 & 2 & 0.6086  & 0.5114  & 0.4910  & 0.7172  & 0.4344  & 0.0456  & -0.0360  & 0.8689  \\
5 & 7 & 2 & 1 & 0 & 1 & -1.5 & -2 & 0.5000  & 0.9207  & 0.2751  & 0.0793  & 0.0000  & 1.6830  & -0.8997  & -1.6830  \\
2 & 8 & 8 & 7 & -1.5 & 1.5 & 1.5 & 1 & 0.1758  & 0.8242  & 0.8242  & 0.8242  & -1.2969  & 1.2969  & 1.2969  & 1.2969  \\
3 & 5 & 1 & 6 & -1 & 0 & -2 & 0.5 & 0.2837  & 0.6065  & 0.0658  & 0.5000  & -0.8653  & 0.4260  & -1.7370  & 0.0000  \\
8 & 2 & 3 & 8 & 1.5 & -1.5 & -1 & 1.5 & 0.8208  & 0.1875  & 0.1792  & 0.9065  & 1.2833  & -1.2501  & -1.2833  & 1.6262  \\
7 & 4 & 9 & 2 & 1 & -0.5 & 2 & -1.5 & 0.7127  & 0.3945  & 0.9214  & 0.1705  & 0.8509  & -0.4220  & 1.6858  & -1.3181  \\
   
   \hline
   \multicolumn{4}{c|}{$U_{16}(16^4)$} & \multicolumn{4}{c|}{$U_{16}(16^4)\in [-2,2]^4$} & \multicolumn{4}{c|}{$U^{new}_{16}(16^4)\in [0,1]^4$} &\multicolumn{4}{c}{$U^{new}_{16}(16^4)\in [-2,2]^4$}\\
   \hline
   1 & 10 & 4 & 6 & -2.00  & 0.40  & -1.20  & -0.67  & 0.4678 & 0.6593 & 0.4717 & 0.7195 & -0.1290  & 0.6370  & -0.1134  & 0.8778  \\
2 & 4 & 13 & 15 & -1.73  & -1.20  & 1.20  & 1.73  & 0.2873 & 0.3469 & 0.0440 & 0.7769 & -0.8510  & -0.6126  & -1.8242  & 1.1074  \\
3 & 13 & 10 & 10 & -1.47  & 1.20  & 0.40  & 0.40  & 0.1599 & 0.7847 & 0.5913 & 0.5975 & -1.3606  & 1.1386  & 0.3650  & 0.3898  \\
4 & 8 & 7 & 1 & -1.20  & -0.13  & -0.40  & -2.00  & 0.0415 & 0.5965 & 0.2213 & 0.3425 & -1.8342  & 0.3858  & -1.1150  & -0.6302  \\
5 & 6 & 1 & 2 & -0.93  & -0.67  & -2.00  & -1.73  & 0.7808 & 0.0927 & 0.3474 & 0.6514 & 1.1230  & -1.6294  & -0.6106  & 0.6054  \\
6 & 15 & 15 & 4 & -0.67  & 1.73  & 1.73  & -1.20  & 0.2165 & 0.4656 & 0.4057 & 0.0389 & -1.1342  & -0.1378  & -0.3774  & -1.8446  \\
7 & 1 & 11 & 7 & -0.40  & -2.00  & 0.67  & -0.40  & 0.7252 & 0.7091 & 0.9508 & 0.0981 & 0.9006  & 0.8362  & 1.8030  & -1.6078  \\
8 & 16 & 8 & 14 & -0.13  & 2.00  & -0.13  & 1.47  & 0.3397 & 0.9060 & 0.7747 & 0.2254 & -0.6414  & 1.6238  & 1.0986  & -1.0986  \\
9 & 3 & 3 & 3 & 0.13  & -1.47  & -1.47  & -1.47  & 0.8519 & 0.5300 & 0.7172 & 0.8348 & 1.4074  & 0.1198  & 0.8686  & 1.3390  \\
10 & 7 & 16 & 9 & 0.40  & -0.40  & 2.00  & 0.13  & 0.6437 & 0.9514 & 0.2860 & 0.9514 & 0.5746  & 1.8054  & -0.8562  & 1.8054  \\
11 & 11 & 5 & 16 & 0.67  & 0.67  & -0.93  & 2.00  & 0.4103 & 0.0373 & 0.6566 & 0.4060 & -0.3590  & -1.8510  & 0.6262  & -0.3762  \\
12 & 12 & 12 & 2 & 0.93  & 0.93  & 0.93  & -1.73  & 0.5944 & 0.4045 & 0.8392 & 0.5308 & 0.3774  & -0.3822  & 1.3566  & 0.1230  \\
13 & 2 & 6 & 11 & 1.20  & -1.73  & -0.67  & 0.67  & 0.9013 & 0.8316 & 0.0971 & 0.4668 & 1.6050  & 1.3262  & -1.6118  & -0.1330  \\
14 & 14 & 2 & 8 & 1.47  & 1.47  & -1.73  & -0.13  & 0.9600 & 0.2836 & 0.5329 & 0.2779 & 1.8398  & -0.8658  & 0.1314  & -0.8886  \\
15 & 9 & 14 & 13 & 1.73  & 0.13  & 1.47  & 1.20  & 0.5341 & 0.2096 & 0.1563 & 0.1535 & 0.1362  & -1.1618  & -1.3750  & -1.3862  \\
16 & 5 & 9 & 5 & 2.00  & -0.93  & 0.13  & -0.93  & 0.1032 & 0.1681 & 0.8969 & 0.8969 & -1.5874  & -1.3278  & 1.5874  & 1.5874  \\
   \hline
   \end{tabular}}
   \label{UDwood}
\end{table}

\begin{table}[h]
   \centering
   \caption{The $\sqrt{MSE}$ of Kriging models in wood model}
   \begin{tabular}{c|rrrr}
   \hline
&\multicolumn{1}{c}{$U_{9}(9^4)$} &\multicolumn{1}{c}{$U^{new}_{9}(9^4)$} & \multicolumn{1}{c}{Avg. $LHS$}&\multicolumn{1}{c}{Avg. $MLHS$}\\
\hline
{\it Poly0} & 957.9618 & 826.0637 & 816.8131 & 805.1325 \\
{\it Poly1} & 878.1982 & 695.0498 & 877.2432 & 847.6947 \\
\hline
   \hline
&\multicolumn{1}{c}{$U_{16}(16^4)$} &\multicolumn{1}{c}{$U^{new}_{16}(16^4)$} & \multicolumn{1}{c}{Avg. $LHS$}&\multicolumn{1}{c}{Avg. $MLHS$}\\
\hline
{\it Poly0} & 829.4668 & 743.3496 & 672.0404 & 725.3436 \\
{\it Poly1} & 779.9872 & 722.6574 & 667.8994 & 765.7914 \\
\hline
   \hline
&\multicolumn{1}{c}{$U_{18}(18^4)$} &\multicolumn{1}{c}{$U^{new}_{18}(18^4)$} & \multicolumn{1}{c}{Avg. $LHS$}&\multicolumn{1}{c}{Avg. $MLHS$}\\
\hline
{\it Poly0} & 691.7101 & 668.6842 & 731.2896 & 685.1813 \\
{\it Poly1} & 618.7809 & 587.3990 & 717.9110 & 711.0604 \\
\hline
   \hline
&\multicolumn{1}{c}{$U_{27}(27^4)$} &\multicolumn{1}{c}{$U^{new}_{27}(27^4)$} & \multicolumn{1}{c}{Avg. $LHS$}&\multicolumn{1}{c}{Avg. $MLHS$}\\
\hline
{\it Poly0} & 2768.7 & 543.0675 & 568.6300 & 558.2930 \\
{\it Poly1} & 1743.9 & 534.7528 & 653.5637 & 603.9952 \\
\hline
   \end{tabular}
   \label{MSEwood}
\end{table}

Besides Table \ref{UDwood}, we also implement 18-run and 27-run designs, LHS, and MLHS. We use the above mentioned designs with {\it Poly0} and {\it Poly1} bases to develop Kriging models. For convenience of reading, $\sqrt{MSE}$ are calculated and listed in Table \ref{MSEwood}. It is obvious that new uniform designs over continuous domain perform better than recorded ones in wood model.

\subsection{\textbf{ \emph{ Six-hump Camelback Model}}}
\begin{figure}[h]
\centering
\includegraphics[width=2.5in]{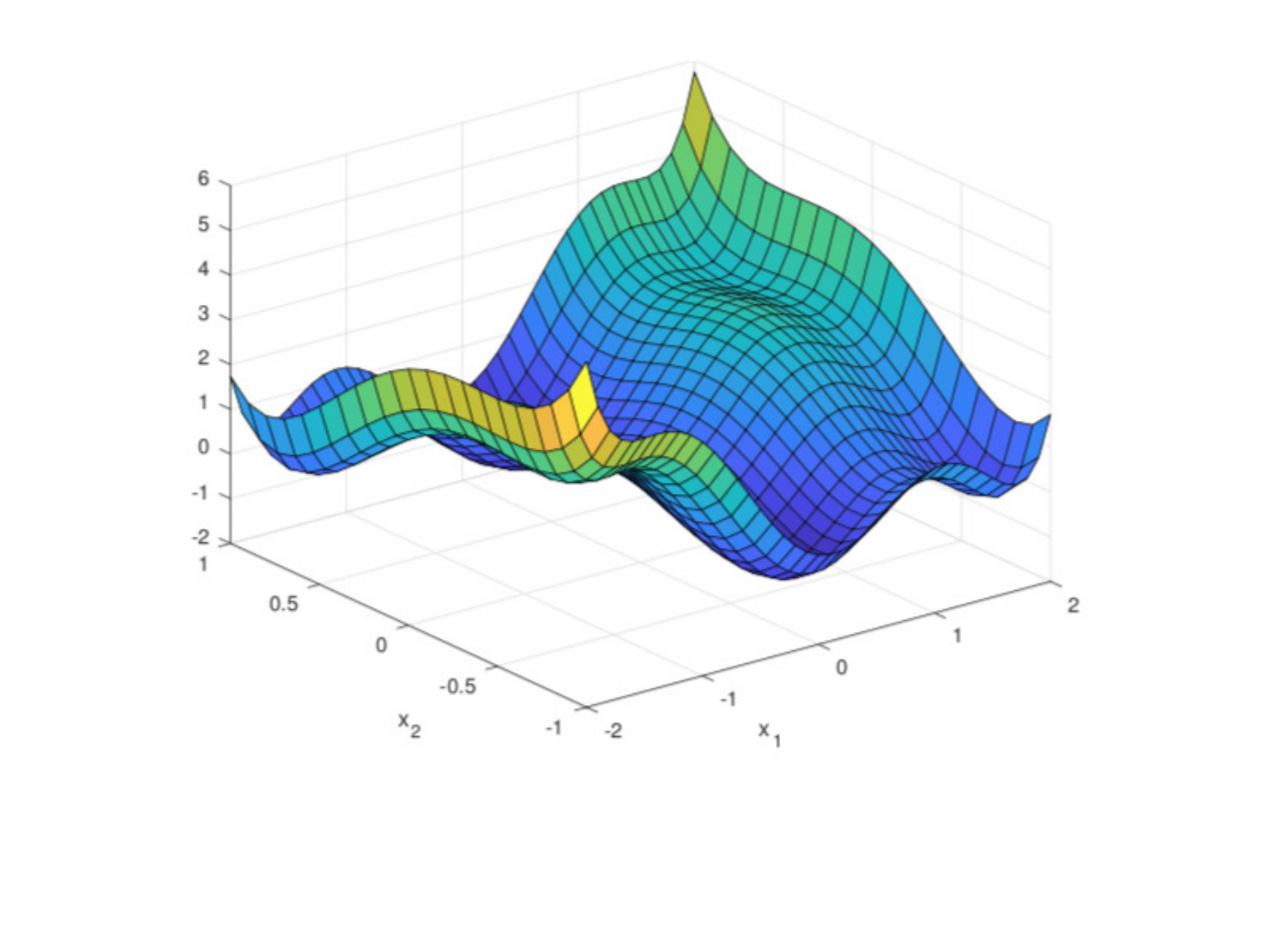}
\caption{The surface of six-hump camelback function} \label{six-hump}
\end{figure}

The function with six-hump camelback surface and six local minima is defined by 
$$Y=4x_1^2-2.1x_1^4+\frac{1}{3}x_1^6+x_1x_2-4x_2^2+4x_2^4,~(x_1,x_2) \in [-3,3]\times[-2,2].$$
The global minimum is $Y^*=-1.03$, when $\x^*=[0.09,-0.71]$ or
$[-0.09,0.71]$. Figure \ref{six-hump} presents the surface of the six-hump Camelback function on $[-2,2]\times[-1,1]$.
 Tabel \ref{MSEcamel} presents the prediction MSE of  Kriging models developed from the 16-run and 27-run designs, LHS and MLHS with {\it Poly0}, {\it Poly1} and {\it Poly2} bases respectively. The new uniform designs, $U^{new}_{16}(16^2)$ and $U^{new}_{27}(27^2)$, perform better in six-hump camelback model as well.

\begin{table}[h]
   \centering
   \caption{The MSE of Kriging models in six-hump camelback model}
   \begin{tabular}{c|rrrr}
   \hline
& \multicolumn{1}{c}{$U_{16}(16^2)$}&\multicolumn{1}{c}{$U^{new}_{16}(16^2)$}&\multicolumn{1}{c}{Avg. $LHS$}&\multicolumn{1}{c}{Avg. $MLHS$}\\
\hline
{\it Poly0} & 316.0265 & 210.8945 & 383.1213 & 338.0651 \\
{\it Poly1} & 316.0265 & 210.8945 & 426.6880 & 380.9689 \\
{\it Poly2} & 253.8087 & 182.3029 & 215.1144 & 221.8797 \\
\hline
   \hline
& \multicolumn{1}{c}{$U_{27}(27^2)$}&\multicolumn{1}{c}{$U^{new}_{27}(27^2)$}&\multicolumn{1}{c}{Avg. $LHS$}&\multicolumn{1}{c}{Avg. $MLHS$}\\
\hline
{\it Poly0} & 159.2230 & 178.3153 & 193.8387 & 176.9656 \\
{\it Poly1} & 159.7014 & 178.9219 & 176.9656 & 182.0625 \\
{\it Poly2} & 78.9778 & 93.0508 & 164.9395 & 115.1672 \\
\hline
   \end{tabular}
   \label{MSEcamel}
\end{table}

\section{Conclusion}
 To yield a new uniform design over continuous experimental domain under the measurement of the centered $L_2$-discrepancy (CD), that is a high dimensional ($p=ns$, where $n$ is the number of runs and
$s$ the number of factors) non-convex optimization problem, we adopt three coordinate descent algorithms with a U-type initialization derived by TA algorithms.
It turns out that the CD values of UDs over continuous domain has been reduced considerably.

\begin{figure} [h]
\centering
\includegraphics[height=0.4\textwidth,width=0.6\textwidth]{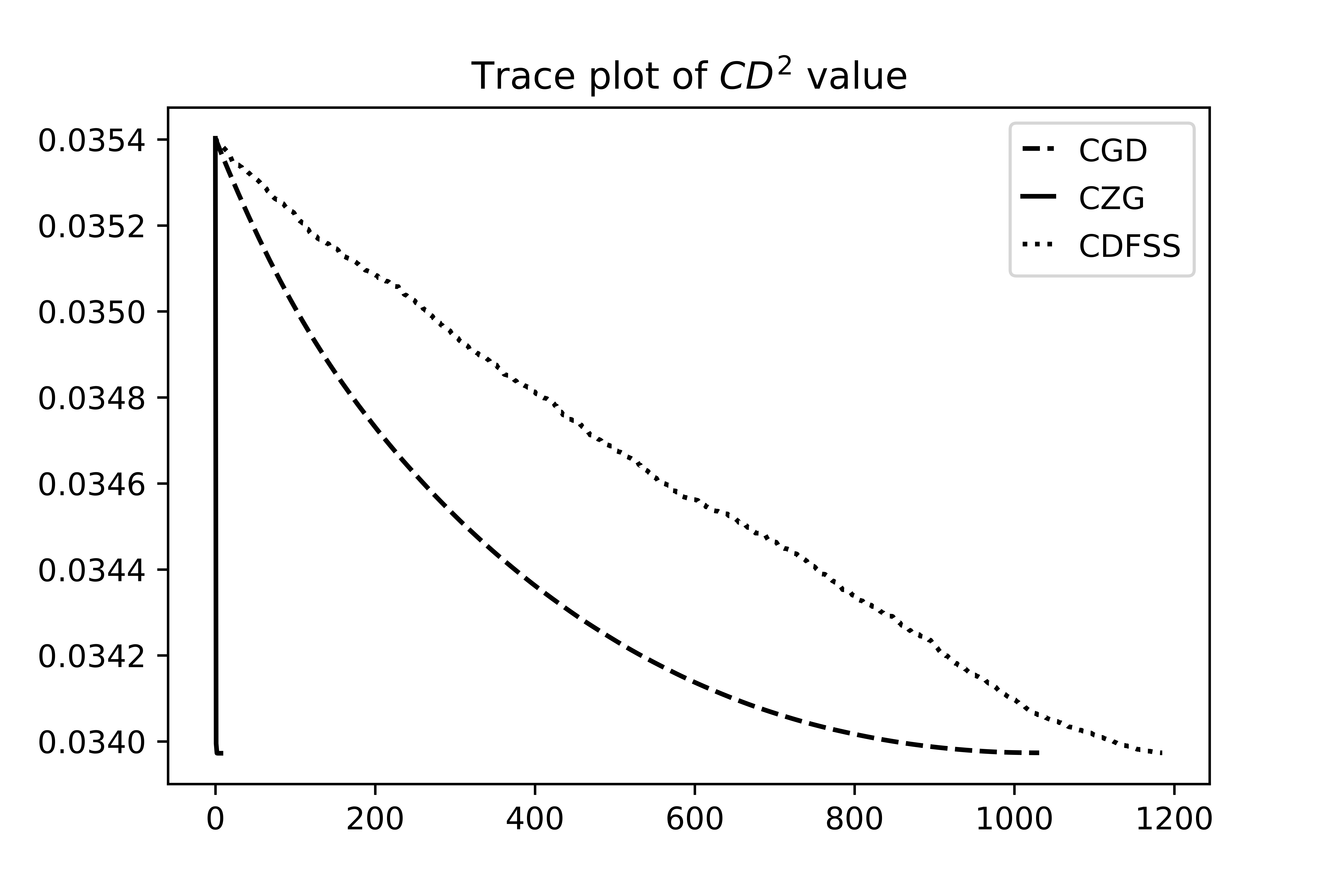}
\caption{$CD^2$ trace plot of three algorithms}
\label{Traceplot}
\end{figure}

The Algorithms 1, 2, and 3  are three coordinate descent variants, performing similarly in our case. In Figure \ref{Traceplot}, we give the trace plot of CD in the above three algorithms. 
Coordinate zero-gradient (Algorithm 2) is the fastest algorithm among them. As for coordinate descent with fixed step size (Algorithm 3), it does not need the
coordinate gradient. 
Similar to gradient descent, coordinate gradient descent
is stable in each step. In addition, coordinate gradient descent adjusts the position of each design point in each dimension.
Compared with gradient descent algorithm, coordinate descent algorithms are
simpler and cheaper. Moreover, each step in coordinate descent is meaningful, which can be visualized during the optimization process.
The visualization is helpful for comprehending uniformity and the optimization process.
Furthermore, it is convenient to be generalized into practical implementation on the experiments with required minimum unit of factors.

Subsequently, to evaluate the new UDs over continuous domain, we compare the modeling performance of the new designs with the recorded UDs on the website,
as well as Latin hypercube sampling in two computer experiments as illustrative instances. The prediction MSE is adopted as evaluation criteria. In this paper, we take Kriging modeling technique to yield the metamodels, that may affect the MSE as well. Nevertheless, it indicates that the new uniform designs are advantaged
on modeling in computer experiments. Uniform designs over continuous domain is promising since designs with continuous factors become feasible and in demand with the improvement of computational ability.

\section{Acknowledgement}
This work was partially supported by the Zhuhai Premier Discipline Grant;  Guangdong Natural Science Foundation under Grant No. 2018A0303130231; and BNU-HKBU United International College under Grant R202108. We thank Dr. Peng Heng for his support.

\bibliographystyle{apalike}
\bibliography{references}
\end{document}